\documentclass[12pt,eqsecnum,preprint]{aastex}
\begin{document}
\title{An Independent Derivation of the Oxford Jet Kinetic Luminosity
Formula}
\author{Brian Punsly}
\affil{4014 Emerald Street No.116, Torrance CA, USA 90503 and
International Center for Relativistic Astrophysics,
I.C.R.A.,University of Rome La Sapienza, I-00185 Roma, Italy}
\email{brian.m.punsly@boeing.com or brian.punsly@gte.net}
\begin{abstract}
This letter presents a theoretical derivation of an estimate for a
radio source jet kinetic luminosity. The expression yields jet
powers that are quantitatively similar to a more sophisticated
empirical relation published by the Willott, Blundell and Rawlings
at Oxford. The formula allows one to estimate the jet kinetic
luminosity from the measurement of the optically thin radio lobe
emission in quasars and radio galaxies. Motivated by recent X-ray
observation, the derivation assumes that most of the energy in the
lobes is in plasma thermal energy with a negligible contribution
from magnetic energy (not equipartition). The close agreement of
the two independent expressions makes the veracity of these
estimates seem very plausible.
\end{abstract}
\section{Introduction}
 The purpose of this
letter is to discuss estimates of the power transported by the
radio jets in quasars and radio galaxies. An accurate estimate of
the jet power is of fundamental physical interest, since it can be
used to quantify the power emerging from the central engine of the
radio source. In actuality, the radio luminosity is merely an
indirect measure of the energy transported through the jets from
the central engine that is not readily interpretable. Most of the
energy flux is in mechanical form (kinetic luminosity) - the
particles and fields necessary to produce the synchrotron
luminosity that is detected in the radio lobes. The radiation
losses, manifested as radio emission from the jet, are merely the
waste energy of this kinematic flow.
\par Surprisingly, the most difficult methods of estimating jet power rely on observations
of the jets themselves. Due to significant Doppler enhancement in
relativistic jets, the synchrotron radio emission is a poor
indicator of intrinsic jet power. For example, Cygnus A has
extremely powerful radio lobes and faint radio jets. Most of the
energy in the jets is not radiated away, but is transported to the
lobes in the classical FRII double lobe morphology. Even the
inclusion of observations of high energy emission such as optical
or X-ray (inverse Compton) in one's analysis of jet energetics
does not tightly constrain the bulk jet flow. If the resolution is
poor at high frequency (as is often the case) then one can not
necessarily associate the plasma emitting the high frequency
photons with the radio emitting plasma. If one has high resolution
images then the high frequency emission can be detected in
enhancement regions or knots in the jets. One can use this
information to get an estimate of the plasma conditions within the
dissipative knot, but this do not necessarily constrain the plasma
state in the bulk of the jet. Furthermore, there are still
ambiguities with the Doppler factor that affect the estimates
quite dramatically.
\par The Doppler enhancement of relativistic flows in jets is a crucial parameter since
the total luminosity of an unresolved jet scales as the Doppler
factor to the fourth power and to the third power for a resolved
cylindrical jet \citep{lin85}. This is the reason why the
implementation of 5 GHz flux densities, as is common in studies of
radio loudness of large quasar samples, is a poor indicator of the
true intrinsic kinetic luminosity of the jets. More specifically,
the majority of core dominated blazar-like quasars have incredibly
strong 5 GHz flux densities from emission on the subkiloparsec
scale, yet they have weak or moderate radio lobe emission
\citet{pun95}. This is interpreted as the jet being of modest
kinetic luminosity (at most) because there is not a large amount
of hot plasma and gas that has been transported through the jets
to the radio lobes. The 5 GHz flux only represents the dissipation
in the jet itself and it has been extremely Doppler boosted. An
estimate of kinetic luminosity based on 5 GHz flux density can be
off by four or more orders of magnitude for a core dominated
blazar.
\par A far better way to estimate the kinetic luminosity from a jet is to study the
isotropic properties of the material ejected from the ends of the
jets in the radio lobes. The radiation from the lobe material is
generally considered to be of low enough bulk velocity so that
Doppler enhancement is not much of an issue. The basic idea is
that lobe expansion is dictated by the internal dynamics of the
lobes and the physical state of the enveloping extragalactic gas.
X-ray observations can indicate a bremsstrahlung spectrum of the
surrounding gas that can be used to find the pressure of the
extragalactic medium. X-rays also provide information on the
working surfaces at the end of the lobes, "the hot spots." One can
associate the X-ray emission as inverse Compton radiation from the
hot spots and the radio luminosity is the synchrotron emission
from the hot spots. This constrains the plasma state within the
luminous hot spots. However, most of the energy stored in the
lobes is in the large diffuse regions of radio emission that
constitutes the majority of the large volume of the radio lobes.
It is the enormous volume of synchrotron emitting plasma within
the lobes ($\sim 10^{4} -10^{5}\, \mathrm{kpc}^{3}$) that is the
most direct indicator that the jets must be supplying huge
quantities of hot plasma and magnetic field energy to the lobes.
One can also use the curvature of the lobe synchrotron radio
spectra to estimate parameters in the diffuse lobe gas - this is
known as spectral ageing. Of course, all of these plasma state
estimations are most accurate when applied to situations in which
one has deep X-ray and radio data of a relaxed classical double
lobe structure. This only occurs in a few instances, so such
detailed analysis are not compatible with large sample studies.
Motivated by these limitations, this letter presents two
techniques for estimating jet energy based on partial information
on the lobe parameters. The two methods involve different
assumptions and have different ambiguities.
\par The most sophisticated calculation of the jet kinetic luminosity
incorporates deviations from the minimum energy estimates in a
multiplicative factor $f$ that represents the small departures
from minimum energy, geometric effects, filling factors, protonic
contributions and low frequency cutoff (see \citet{wil99} for
details). The quantity, $f$, is argued to constrained to be
between 1 and 20. In \citet{blu00}, it was further determined that
$f$ is most likely in the range of 10 to 20. Therefore we choose
$f=15$ in order to convert 151 MHz flux densities, $F_{151}$, to
estimates of kinetic luminosity, $Q_{151}$, using equation (12)
and figure 7 of \citet{wil99},
\begin{eqnarray}
&& Q_{151}\approx 1.1\times
10^{45}\left[(1+z)^{1+\alpha}Z^{2}F_{151}\right]^{\frac{6}{7}}\mathrm{ergs/sec}\;,\\
&& Z \approx 3.31-3.65\times \nonumber\\
 &&
\left(\left[(1+z)^{4}-0.203(1+z)^{3}+0.749(1+z)^{2}+0.444(1+z)+0.205\right]^{-0.125}\right)\;.
\end{eqnarray}
The quantity $F_{151}$ is the optically thin flux density from the
lobes (i.e., no contribution from Doppler boosted jets or radio
cores) measured at 151 MHz in Jy. The flux density spectral index
is defined as $F_{\nu}\sim\nu^{-\alpha}$. We have assumed a
cosmology with $H_{0}$=70 km/s/Mpc, $\Omega_{\Lambda}=0.7$ and
$\Omega_{m}=0.3$. The expression for Z is from \citet{pen99}.
\par In the following, a new estimate of the jet kinetic
luminosity is derived that is motivated by the wealth of X-ray
data on radio lobes that has been published since
\citet{wil99,blu00}. Both the current manuscript and the
\citet{wil99} derivations rest on the basic premise that $Q = U/T$
+ radiation losses, where $U$ is the energy stored in the lobes
and $T$ is the elapsed time. In \citet{wil99}, $U$ is found by
assuming the lobes are near equipartition and there is uncertainty
in the energy from protonic components and low frequency cutoffs
which are incorporated in the empirical factor $f$ discussed
above. Conversely, motivated by the new X-ray data described
below, $U$ is computed theoretically in the limit that the lobes
are very far from equipartition. In \citet{wil99}, $T$ is
determined by an empirical estimator for the age of the radio
source based on their lengths and head advance speeds. Conversely,
in this treatment $T$ is computed from spectral ageing. In spite
of the fact that $U$ and $T$ are determined from completely
different methods and assumptions, the expressions for $Q$ that
are found in (1.1) and (3.9) yield similar values (to within a
factor of 2) for the jet luminosity. This close agreement lends
credence to the claim that these formulae are robust estimators of
jet kinetic luminosity.
\section{Motivation: X-ray Observations}The minimum energy condition in the lobes seems to be in conflict
with the X-ray data on the surrounding extragalactic gas. This was
first noted for Cygnus A (see \citet{pun01} and references
therein) and later for 3C 388 in \citet{lea01} as well as for a
large sample of FRII radio sources in \citet{har00} based on ROSAT
data. These studies concluded that typically the pressure in the
external gas greatly exceeds (by at least an order of magnitude)
the lobe pressure associated with the minimum energy assumption.
The general picture that seems to be emerging from the X-ray data
of ROSAT, ASCA, XMM and Chandra is that the hot spot energies seem
to slightly exceed the minimum energy requirement based on inverse
Compton calculations of X-rays from the hot spots (see for example
\citet{wil02,brn02}), but the lobes themselves are far from
equipartition. In \citet{lea01}, it was deduced that the departure
from equipartion was most likely the consequence of a low energy
population of positrons and electrons that is not an extension of
the power law distribution responsible for the radio emission, but
a low energy excess. The other possibility is protonic matter
which would also drastically increase the energy content of the
lobes over the equipartition estimates. In either case, the field
energy is only a few percent of the particle energy in the lobes.
For example, high resolution Chandra data was used to model the
X-rays as inverse Compton emission in the lobes in the radio
galaxy 3C 219 \citep{bru02}. It was concluded that the particle
energy exceeded the magnetic energy by a factor of 60 in the radio
lobes. Similarly, the FRII radio galaxy 3C 452 was studied with
Chandra in \citet{iso02} who estimate that the energy density in
the particles is 27 times that of the magnetic field in the lobes.
For most FR II sources, typical magnetic field strengths in the
lobes that are estimated from X-ray data and pressure balance are
only a third to a fifth of the minimum energy value
\citep{har00,lea01}.
\par These X-ray observations yield valuable information on the energy
content of the lobes, but are disjoint from the spectral ageing
estimates of the lobe advance speeds. In the following estimation
of jet kinetic energy, it is assumed based on the X-ray data
presented above that the energy content of the lobes is purely in
particle form to first order (accurate to a few percent). Yet, the
notion that spectral ageing provides an estimate of lobe age is
retained. By choosing a subequipartion field strength in the
lobes, spectral ageing estimates are found to be longer than the
corresponding minimum energy estimates \citep{ale96}. Similarly,
by setting $t_{sep}=t_{syn}$, the subequipartion fields yield
lower lobe separation velocities if this spectral ageing argument
is viable. Thus, the problem of the large lobe advanced speeds in
the minimum energy assumption is remedied by this modification
\citep{ale96}. This method of computing jet kinetic luminosity is
the lowest order improvement to the minimum energy estimate and
was implemented in \citet{pun01} to study Cygnus A.
\section{Particle Energy Dominated Lobes} Motivated by the X-ray observations,
 we proceed to compute
the jet power based on the limit that all of the lobe energy is in
the hot particles. We also assume that the time to convert a jet
energy flux to the stored lobe energy is the time that it has
taken the lobes to propagate from the central engine to their
current separation, $t_{sep}$.
\subsection{Spectral Ageing}Spectral ageing within the radio lobes
is often used to determine the lobe plasma age.
 The results are predicated on the assumption that the lobe plasma is primarily back flowing
 plasma in the sense that jet plasma is deflected backward at the working surfaces in the
 hot spots to form the lobe plasma. By studying the curvature of the radio spectra at
 different points within the lobes, one can in principle (if there is no reheating or
 re-injection of the plasma) determine the gradient in the high energy cutoff of the
 electron distribution due to synchrotron cooling and hence the plasma age. The age
 of the lobe plasma closest to the central engine should be the oldest plasma. Thus,
 one has an estimate of lobe age and therefore the lobe advance speed. Defining the spectral break frequency as $\nu_{_{b}}$, the
synchrotron lifetime is expressed in cgs units as \citep{liu92}
\begin{eqnarray}
&& t_{syn}\approx 1.58\times 10^{12}
B^{-\frac{3}{2}}\nu_{_{b}}^{-\frac{1}{2}}\;.
\end{eqnarray}
\subsection{The Energy Contained Within the Synchrotron Emitting
Plasma} Consider a power law distribution of energetic particles
(probably electrons and positrons) expressed in terms of the
thermal Lorentz factor, $\gamma$, for a uniform source in a
volume, $V$. The total number of particles contributing to the
synchrotron radiation in the frequency interval $\nu_{_1}\leq\nu
\leq \nu_{_2}$ is:
\begin{eqnarray}
&&
N_{r}=N_{0}V\int_{\gamma_{_1}}^{\gamma_{_2}}\gamma^{-n}\,\mathrm{d}\gamma
\; .\end{eqnarray} The minimum and maximum Lorentz factors in the
expression above are related to lower and upper cutoff frequencies
in the synchrotron spectrum $\nu_{_{1}}$ and $\nu_{_{2}}$,
respectively by \citep {gin79}
\begin{eqnarray}
&&
\gamma_{_1}=\left[\frac{2\nu_{_1}y_{_1}(n)}{3\nu_{_B}}\right]^{\frac{1}{2}}\;,\quad
\gamma_{_2}=\left[\frac{2\nu_{_2}y_{_2}(n)}{3\nu_{_B}}\right]^{\frac{1}{2}},
\end{eqnarray}
where $\nu_{_{B}}=(eB)/(2\pi{m_{e}c})$ is the cyclotron frequency
and note that
\begin{eqnarray}
&& y_{_1}(n) = 2.2,\, y_{_2}(n)=0.10 ,\mathrm{if}\;
n=2.5:\mathrm{and}\, y_{_1}(n) = 2.7 ,\,
y_{_2}(n)=0.18,\mathrm{if}\; n=3.0\;.
\end{eqnarray}
The synchrotron spectral luminosity of the plasma, $L(\nu)$, is a
function of both the particle distribution in momentum space and
the magnetic field strength. Integrating the synchrotron power
formula over the particle distribution yields \citep{gin79}
\begin{eqnarray}
&& U_{e}\approx\frac{2\times 10^{11}B^{-\frac{3}{2}}}
{a(n)(n-2)}L(\nu_{_1})\nu_{_1}^{\frac{1}{2}}
[y_{_1}(n)]^{\frac{n-1}{2}}\nonumber \\
&& \hspace{0.3in} \times
\left[1-\left(\frac{y_{_2}(n)\nu_{_1}}{y_{_1}(n)\nu_{_2}}\right)^{\frac{n-1}{2}}\right
]\;,
\end{eqnarray}
where
\begin{eqnarray}
&& a(n)=\frac{\left(2^{\frac{n-1}{2}}\sqrt{3}\right)
\Gamma\left(\frac{3n-1}{12}\right)\Gamma\left(\frac{3n+19}{12}\right)
\Gamma\left(\frac{n+5}{4}\right)}
{8\sqrt\pi(n+1)\Gamma\left(\frac{n+7}{4}\right)} \; .
\end{eqnarray}
\subsection{Estimating the Jet Power}
Set $t_{sep}$ equal to the synchrotron ageing timescale $t_{syn}$,
associated with the spectral break in the flux density, $F_{\nu}$,
of the lobe plasma closest to the quasar (the emission just above
the spectral break is from the lowest energy electrons that have
synchrotron radiated away their energy and hence the oldest
subpopulation of charges that have experienced synchrotron decay
in the lobes). By combining $t_{syn}$ from (3.1) with the
expression for the plasma energy, (3.5), one obtains an estimate
for the energy stored in the lobes as a function of spectral
luminosity, $L(\nu)$, in the limit of particle energy dominance,
\begin{eqnarray}
&& U_{e}\approx
\frac{L(\nu_{1})(\nu_{1})^{1/2}(\nu_{b})^{1/2}}{7.9(n-2)a(n)}\left[y_{1}(n)\right]^{\frac{n-1}{2}}t_{syn}\;.
\end{eqnarray}
\par Evaluation of the formula above requires numerous
characteristic frequencies that need to be determined. Since
expressions that are applicable to sparse data are desired for
evaluating large samples, we choose a common set of ``typical''
parameters for an FRII radio source. First of all, determining
$\nu_{b}$ requires high resolution maps of the lobes at a variety
of frequencies. This data has been obtained for only a limited
number of bright sources. The largest sample of these detailed
observations is from \citet{liu92}. The average rest frame break
frequency from the sample of \citet{liu92} is $\nu_{b}=8.9\pm7.0$
GHz. Secondly, in order to estimate the minimum synchrotron
frequency we note that from \citet{bra69} (even though the
measurements are likely to be extremely inaccurate) it is clear
that many FRII sources are very strong emitters down to
frequencies at least as low as 12.6 MHz. Thus, we pick
$\nu_{1}=10\,\mathrm{MHz}$ in the quasar rest frame. Finally, in
order to approximate the total radio luminosity, $L\equiv\int
L(\nu)\,d\nu$ (including the significant contribution at
frequencies above the spectral break), with a single spectral
index, a value of $\nu_{2}=100\,\mathrm{GHz}$ is chosen. Inserting
these ``typical'' frequency values into (3.7), one obtains a
simple estimator of lobe power in the limit of particle dominance,
and noting that at the spectral break frequency, $t_{syn}\approx
t_{sep}$
\begin{eqnarray}
&& Q\approx\frac{U_{e}}{t_{sep}}+L\approx
\frac{\left[y_{1}(n)\right]^{\frac{n-1}{2}}(15.1)^{\alpha}}{(n-2)a(n)}10^{42}(1+z)^{1+\alpha}Z^{2}F_{151}\,\mathrm{ergs/sec}+L\;,
\end{eqnarray}
where the spectral index $\alpha=(n-1)/2$ has been introduced
($L(\nu)\sim\nu^{-\alpha}$). It should be noted that the estimates
above are very conservative. The existence of a substantial proton
component to the lobe gas or an extension of the low frequency
portion of the electron spectrum would increase the energy flux
estimates significantly. Observations suggest that $\alpha\approx
1$ is a good fiducial value for the expression (3.8)\citep{kel69},
\begin{eqnarray}
&&Q_{par}\approx
5.7\times10^{44}(1+z)^{1+\alpha}Z^{2}F_{151}\,\mathrm{ergs/sec}\;,\quad\alpha\approx
1\;.
\end{eqnarray}
\section{Conclusion}
An independent formula for the jet kinetic luminosity estimator in
(1.1) is derived in (3.9) that was motivated by different physical
assumptions. The two estimates agree to within a factor of 2. This
lends credence to the idea that (1.1) and (3.9) are robust
estimators of jet kinetic luminosity when the optically thin
extended emission is measured in a deep radio map. The main result
of this paper is that (1.1), although very ambitious in its
intent, is likely to be correct to within a factor of a few even
if some of the assumptions in its derivation are inaccurate.

\end{document}